\documentclass[reprint,superscriptaddress,amsmath,amssymb,aps,prd]{revtex4-1}

\usepackage{graphicx}
\usepackage{dcolumn}
\usepackage{bm}
\usepackage{comment}
\usepackage{multirow}
\usepackage{xcolor}
\usepackage[caption=false]{subfig}
\usepackage{soul}

\begin{document}

\title{Tunable photonic crystal haloscope for high-mass axion searches}

\author{Sungjae Bae}
\affiliation{Department of Physics, KAIST, Daejeon 34141, Republic of Korea}
\affiliation{Center for Axion and Precision Physics Research, IBS, Daejeon 34051, Republic of Korea}
\author{SungWoo Youn}
\email{swyoun@ibs.re.kr}
\affiliation{Center for Axion and Precision Physics Research, IBS, Daejeon 34051, Republic of Korea}
\author{Junu Jeong}
\affiliation{Center for Axion and Precision Physics Research, IBS, Daejeon 34051, Republic of Korea}

\date{\today}

\begin{abstract}
In the search for axion dark matter, the cavity-based haloscope offers the most sensitive approach to the theoretically interesting models in the microwave region.
However, experimental searches have been limited to relatively low masses up to a few tens of $\mu$eV, benefiting from large detection volumes and high quality factors for a given experimental setup.
We propose a new cavity design suitable for axion searches in higher mass regions with enhanced performance.
The design features a periodic arrangement of dielectric material in a conventional conducting cavity where the resonant frequency is determined by the interspace.
This photonic crystal haloscope can make full use of a given volume even at high frequencies while substantially improving the cavity quality factor.
An auxetic structure is considered to deploy the array for two-dimensional frequency tuning.
We present the characteristics of this haloscope design and demonstrate its feasibility for high-mass axion searches.
\end{abstract}

\maketitle

\section{Introduction}

The axion is a hypothetical elementary particle, proposed to offer a dynamic solution to the charge-parity problem in the strong interaction of particle physics~\cite{paper:axion1, paper:axion2}.
It has drawn increasing attention as a favored candidate for dark matter~\cite{paper:CDM}. 
The cavity haloscope, a promising detection technique in the microwave region, utilizes a microwave cavity immersed in a strong magnetic field~\cite{paper:sikivie}. 
The axion-photon conversion rate can be resonantly enhanced when the cavity resonant frequency is tuned to the converted photon frequency~\cite{paper:detection_rate}:
\begin{equation}
    P_{a \gamma \gamma } = g_{a\gamma \gamma}^{2} \frac{\rho_a}{m_a} B_{0}^{2} V C \frac{Q_c Q_a}{Q_c + Q_a},
\label{eq:axion-photon_conversion}
\end{equation}
where $g_{a\gamma\gamma}$ is the axion-photon coupling constant, $\rho_a$ is the axion dark matter density, $m_a$ is the axion mass, $B_0$ is the external magnetic field, $V$ is the cavity volume, and $Q_c$ and $Q_a$ are the cavity and axion quality factors. 
The form factor $C$ is a mode dependent parameter, defined as
\begin{equation}
C = \frac{ |\int_{V} E_c \cdot B_0 d^3x |^2} {\int_{V} |B_0|^2 d^3x \int_{V} \epsilon|E_c|^2 d^3x },
\label{eq:form_factor}
\end{equation}
where $E_{c}$ is the electric field of the cavity resonant mode and $\epsilon$ is the dielectric constant inside the cavity. 
Cavity haloscope experiments commonly use a solenoid for the external magnetic field to store the magnetic energy effectively.
For this geometry, the TM$_{010}$ mode of a cylindrical cavity is typically chosen because it yields the highest form factor ($C_{\rm TM_{010}} = 0.69$). 
The axion mass is presently unknown and the possible mass spans many orders of magnitude.
Thus the search strategy relies on how quickly a vast range of masses can be scanned.
The relevant quantity is called the scan rate, which is formulated in terms of experimental parameters as
\begin{equation}
    \frac{d\nu}{dt} \propto B_0^4 T_s^{-2} V^2 C^2 \frac{Q_c Q_a}{Q_c + Q_a},
\label{eq:scan_rate}
\end{equation}
where $T_s$ denotes the system noise temperature.

The resonant frequency of the TM$_{010}$ mode is inversely proportional to the cavity radius, directly affecting the detection volume and quality factor and thus limiting the experimental sensitivity at high frequencies.
Several efforts have been made to address such issues, e.g., arrays of multiple cavities~\cite{thesis:ADMX_multicav, paper:multiple_cavity}, cavities with multiple cells~\cite{paper:CAPP-9T, paper:RADES}, higher-order resonant modes~\cite{paper:supermode, paper:wheel, paper:DBAS}, and high-$Q$ cavities~\cite{paper:SC_QUAX, paper:highQ_QUAX, paper:SC_CAPP}. 
Despite those efforts, the practically accessible frequency regions remain below 10\,GHz.

Meanwhile, new search strategies were proposed by employing a periodic structure of dielectric or metallic materials.
The former exploits the boosting effect arising from constructive interference of the electromagnetic waves generated by the discontinuity of the axion-induced field on the surface of high-$\epsilon$ dielectric disks precisely aligned in one-dimension~\cite{paper:MADMAX}.
The latter, on the other hand, utilizes the resonant effect at the plasma frequency of thin conducting wires in a two-dimensional (2-D) lattice structure~\cite{paper:plasma_haloscope}.
For either scheme, the search frequency is determined by the distance between adjacent disks or wires and is not subject to the physical size of the detector.
This allows for large conversion volumes at high frequencies.

The dielectric boosting system requires a new large-scale experiment to realize the concept with reasonable sensitivity, while the plasma resonator can be implemented in existing cavity experiments.
For the plasma haloscope, in addition, the meta-structure of thin wires features a flat field distribution over the volume, thereby yielding a near-unity form factor. 
However, a large number of conducting thin wires would lead to substantial degradation of the quality factor and create some challenges in cavity construction and frequency tuning.
In this report, we propose a high-performance photonic crystal haloscope design, consisting of a 2-D lattice structure of dielectric rods, and a practical frequency tuning mechanism, the combination of which offers a more effective approach for axion searches beyond $10$\,GHz.

\section{Photonic crystal cavity}

Motivated by the plasma haloscope scheme, we considered a cavity design of periodically placed dielectric rods instead of thin conducting wires.
While retaining the advantages of the plasma haloscope (i.e., large detection volume at high frequencies and resonant frequency tunable by varying the distance between neighboring base materials), a periodic array of dielectric material enables high-quality, low-loss cavity designs.

\begin{figure}
    \centering
    \includegraphics[width=0.9\linewidth]{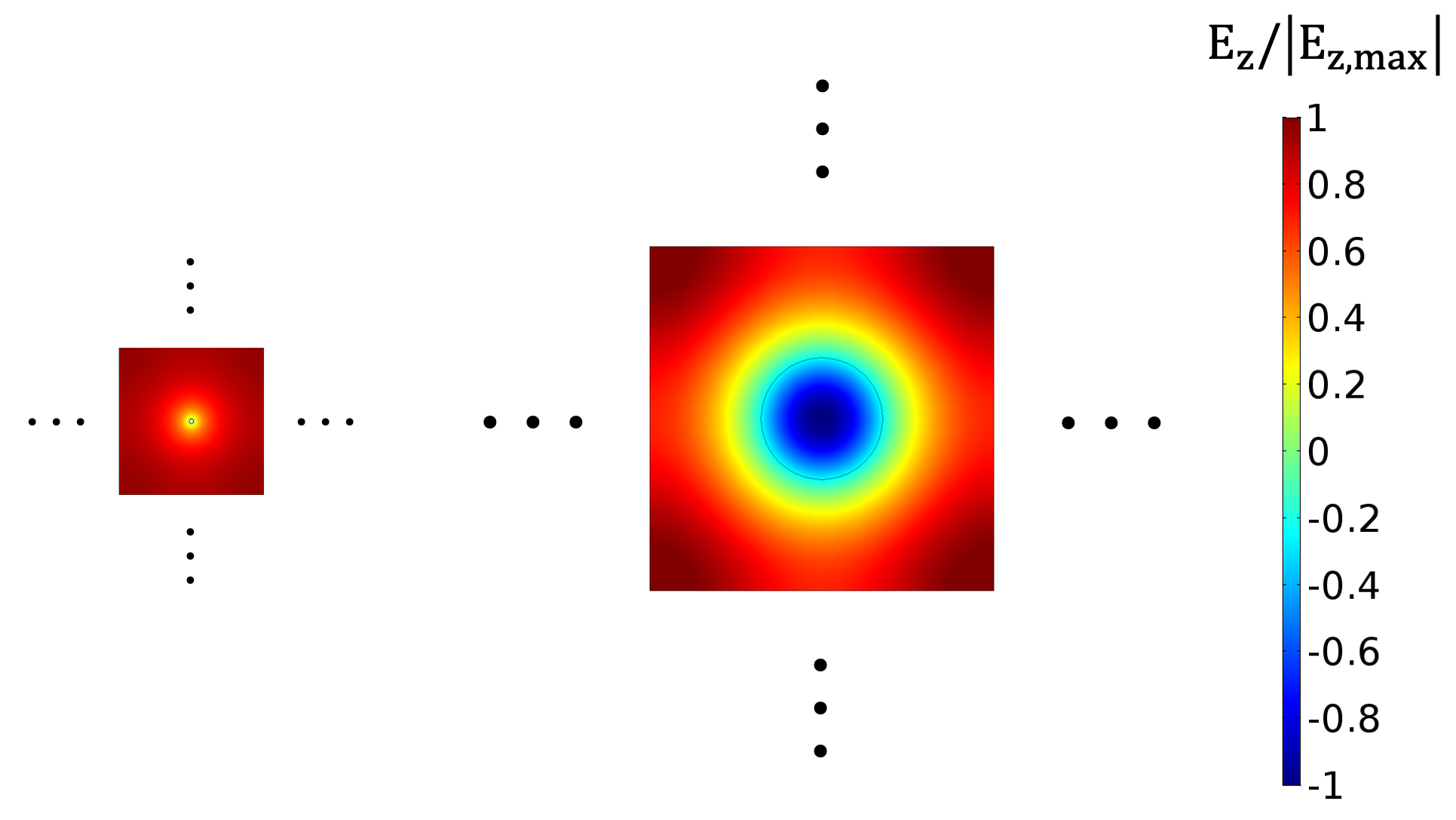}
    \caption{Electric field distributions of square unit cells with a conducting wire (left) and a dielectric rod (right) positioned in the middle. 
    A 2-D photonic crystal can be simulated as a repeating pattern of unit cells with periodic boundary conditions imposed.}
    \label{fig:wire_vs_dielectric}
\end{figure}

The electromagnetic properties of such photonic crystals were assessed based on 2-D simulations using the finite element method~\cite{tool:comsol}.
A lattice structure can be modeled by considering a polygon unit cell with the base material centered, and then arranging the unit cells in a repeating pattern with periodic boundary conditions imposed.
Examples of a square unit cell with a wire and a dielectric are shown in Fig.~\ref{fig:wire_vs_dielectric}, where the dimensions of the cell and base material are determined to produce the same resonant frequency.
From a unit cell point of view, the resonant modes resemble the TM$_{010}$ and TM$_{020}$ modes, respectively.
Strictly speaking, for the dielectric unit cell, our desired mode corresponds to the monopole mode in photonic crystals~\cite{paper:Dirac_cone}.
The monopole TM mode induces a field component that oscillates in the opposite direction inside the dielectric, resulting in a relatively low form factor.
However, dielectric properties substantially improves the quality factor compared to conductors.
Furthermore, the size of the unit cell with dielectrics is larger than that with wires for the same frequency. 
This notably reduces the number density of the base material (e.g., by factors of $>10$ for 10\,GHz and $>5$ for 20\,GHz), which facilitates cavity assembly and frequency tuning.
A low-density base material is also beneficial for reducing the mode population particularly for dielectric which induces additional higher-order modes due to field confinement in it. 
Based on a toy model study where two square cavities filled with the unit cells in Fig.~\ref{fig:wire_vs_dielectric} were simulated, we found that the mode population for a given frequency range is about 10\% smaller for dielectrics compared to conductors.

\begin{figure}
    \centering
    \includegraphics[width=0.9\linewidth]{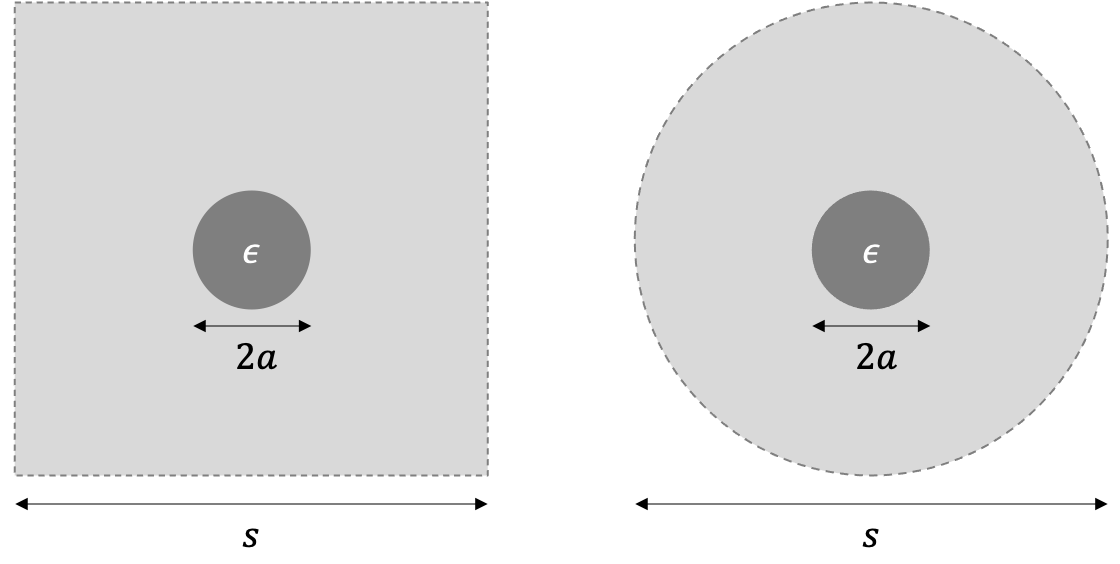}
    \caption{Two-dimensional geometries of dielectric metamaterial unit cell.
    A square shape (left) can be approximated to a circular one (right) for analytical calculation.}
    \label{fig:geometry_2d}
\end{figure}

The analytic solutions of the electromagnetic fields for the photonic crystal can be obtained through geometric approximation.
This involves replacing the square geometry of the unit cell, for which the exact solution is difficult to obtain, with a circular one to take advantage of azimuthal symmetry, as shown in Fig.~\ref{fig:geometry_2d}.
Assuming a circular cell of diameter $s$ with a dielectric rod of radius $a$ and relative permittivity $\epsilon$, Maxwell's equations are given in a cylindrical coordinate system as
\begin{equation}
\label{eq:maxwell_2d}
    \begin{split}
        \partial_{\rho}^{2}A_{z} + \rho^{-1}\partial_{\rho}A_{z} + \omega^{2}A_{z} &= 0 \,\,{\rm for}\,\, \rho > a, \\
        \partial_{\rho}^{2}A_{z} + \rho^{-1}\partial_{\rho}A_{z} + \epsilon \omega^{2}A_{z} &= 0 \,\,{\rm for}\,\, \rho < a,
    \end{split}
\end{equation}
where $A_z$ is the $z$ component of the vector potential and $\omega(=2\pi\nu)$ is the angular frequency.
The boundary conditions to be satisfied are 
\begin{equation}
\label{eq:boundary_2d}
\begin{split}
    \lim_{\rho\to a-}A_{z}(\rho) &= \lim_{\rho\to a+}A_{z}(\rho), \\
    \lim_{\rho\to a-}\partial_{\rho}A_{z}(\rho) &= \lim_{\rho\to a+}\partial_{\rho}A_{z}(\rho), \\
    \lim_{\rho\to s/2-}\partial_{\rho}A_{z}(\rho) &= 0,
\end{split}
\end{equation}
where the $+/-$ sign indicates the inward/outward direction. 
The general solutions of Eq.~\ref{eq:maxwell_2d} for our TM mode are expressed in terms of the cylindrical harmonics as
\begin{widetext}
\begin{equation}
    A_{z}(\rho) = \mathcal{A}
    \begin{cases}
        J_{0}(\sqrt{\epsilon} \omega a) \left[ Y_{1}(\omega s / 2)J_{0}(\omega \rho) - J_{1}(\omega s / 2)Y_{0}(\omega \rho) \right] & {\rm for\,\,} \rho > a, \\
        J_{0}(\sqrt{\epsilon} \omega \rho) \left[Y_{1}(\omega s / 2)J_{0}(\omega a) - J_{1}(\omega s / 2)Y_{0}(\omega a) \right] & {\rm for\,\,} \rho < a,
    \end{cases}
\end{equation}
\end{widetext}
where $J_{\alpha}$ and $Y_{\alpha}$ are the Bessel functions of the first and second kinds of order $\alpha$.
The amplitude $\mathcal{A}$ is determined by the boundary condition of
\begin{equation}
\label{eq:omega_2d}
\begin{split}
    & J_{0}(\sqrt{\epsilon}\omega a)\left[Y_{1}(\omega s / 2)J_{1}(\omega a) - J_{1}(\omega s / 2)Y_{1}(\omega a) \right] = \\
    & \sqrt{\epsilon}J_{1}(\sqrt{\epsilon}\omega a) \left[Y_{1}(\omega s / 2)J_{0}(\omega a) - J_{1}(\omega s / 2)Y_{0}(\omega a) \right].
\end{split}
\end{equation}

\begin{figure}
    \centering
    \includegraphics[width=\linewidth]{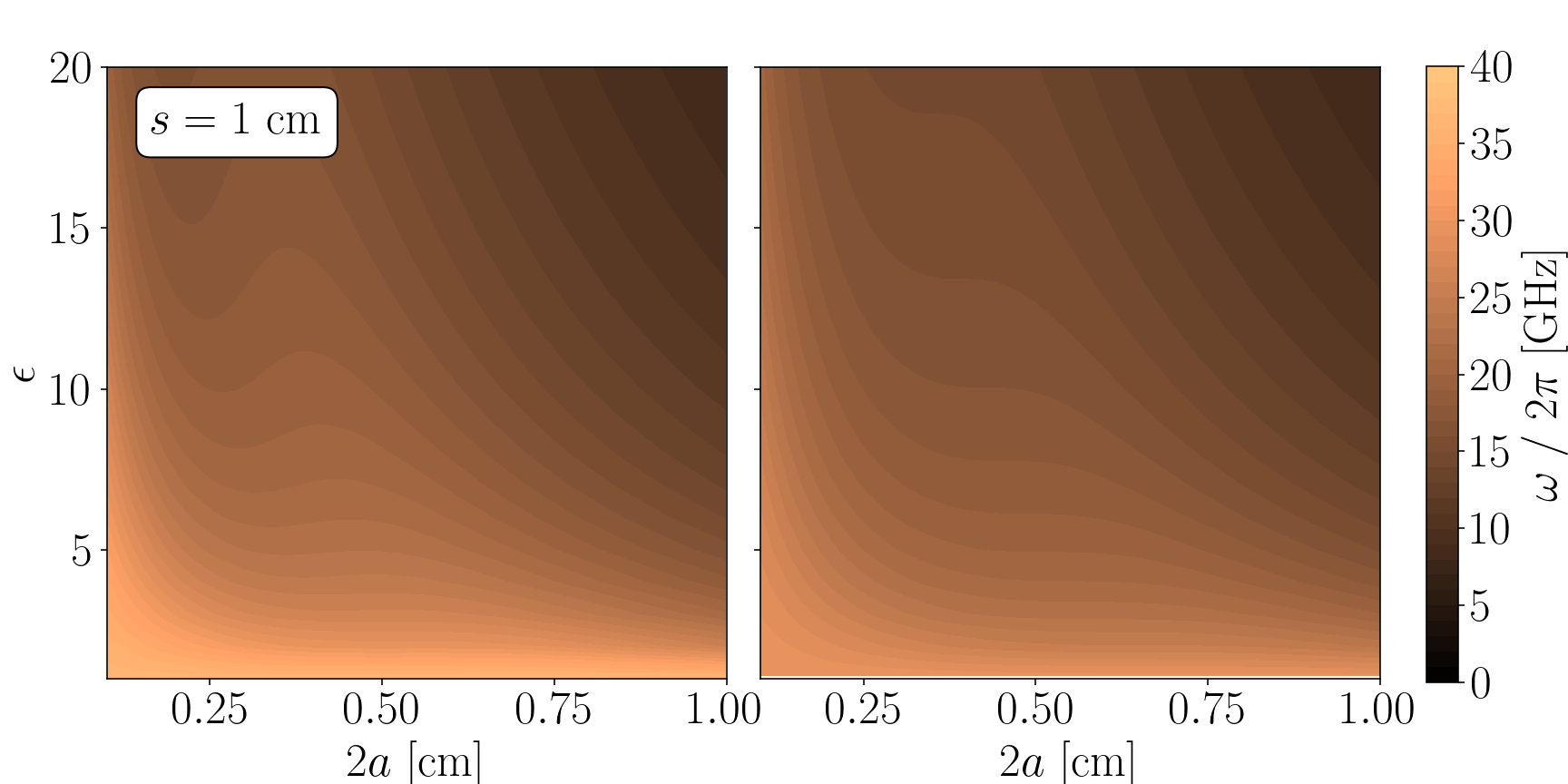} 
    \includegraphics[width=0.68\linewidth]{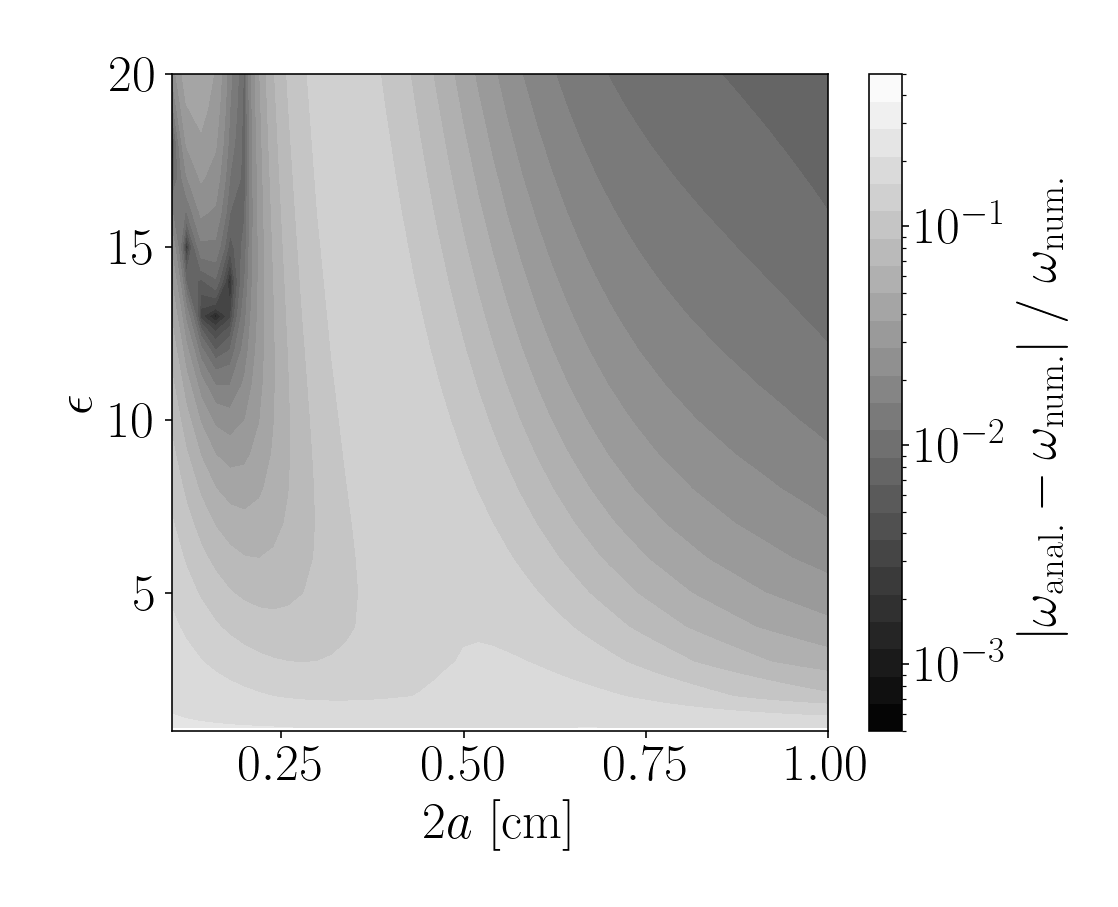}
    \caption{Resonant frequency $\omega$ of our TM mode obtained from the analytical approach (Eq.~\ref{eq:omega_2d_approx}) for the circular geometry (top left) and from a numerical calculation for the square geometry (top right). 
    The lattice size of $s=1$\,cm is assumed.
    The bottom plot shows the difference between the two approaches normalized to the latter.}
    \label{fig:omega_2d}
\end{figure}

Eq.~\ref{eq:omega_2d} is a combination of transcendental functions that are difficult to solve exactly.
With an additional linear approximation, however, it can be transformed into an algebraic equation of $\omega$ by means of the Fourier series expansion.
By defining the lattice constant of this geometric unit cell by $\tilde{s} = (s - 2 a) + 2 \sqrt{\epsilon} a$, the first-order approximation at $\omega = 2\pi/\tilde{s}$ gives a solution for $\omega$,
\begin{equation}
\label{eq:omega_2d_approx}
    \omega \approx \frac{2 \chi}{\tilde{s}} \frac{A_{1} J_{1}(\theta_{1}) + A_{2} J_{2}(\theta_{1}) + B_{1} Y_{1}(\theta_{1}) + B_{2} Y_{2}(\theta_{1})}{C_{1} J_{1}(\theta_{1}) + C_{2} J_{2}(\theta_{1}) + D_{1} Y_{1}(\theta_{1}) + D_{2} Y_{2}(\theta_{1})},
\end{equation}
where $\theta_{1} \equiv \chi s/\tilde{s}$ with $\chi\approx 3.8317$, the first root of $J_1$.
The coefficients are given as
\begin{equation}
\begin{split}
    A_{1} &= {\tilde{s}} J_{0}(\sqrt{\epsilon}\theta_{2})Y_{1}(\theta_{2}) - {2a} \chi J_{0}(\sqrt{\epsilon}\theta_{2})Y_{2}(\theta_{2}) \\
    &\;\;- \sqrt{\epsilon} {\tilde{s}} J_{1}(\sqrt{\epsilon}\theta_{2})Y_{0}(\theta_{2}) + {2 \epsilon a} \chi J_{2}(\sqrt{\epsilon}\theta_{2})Y_{0}(\theta_{2}), \\
    A_{2} &= -{s} \chi J_{0}(\sqrt{\epsilon}\theta_{2})Y_{1}(\theta_{2}) + \sqrt{\epsilon}{s} \chi J_{1}(\sqrt{\epsilon}\theta_{2})Y_{0}(\theta_{2}), \\
    B_{1} &= -{\tilde{s}} J_{0}(\sqrt{\epsilon}\theta_{2})J_{1}(\theta_{2}) + {2a} \chi J_{0}(\sqrt{\epsilon}\theta_{2})J_{2}(\theta_{2}) \\
    &\;\;+ \sqrt{\epsilon} {\tilde{s}} J_{1}(\sqrt{\epsilon}\theta_{2})J_{0}(\theta_{2}) - {2 \epsilon a} \chi J_{2}(\sqrt{\epsilon}\theta_{2})J_{0}(\theta_{2}), \\
    B_{2} &= {s} \chi J_{0}(\sqrt{\epsilon}\theta_{2})J_{1}(\theta_{2}) - \sqrt{\epsilon}{s} \chi J_{1}(\sqrt{\epsilon}\theta_{2})J_{0}(\theta_{2}), \\
    C_{1} &= 2 {\tilde{s}} J_{0}(\sqrt{\epsilon}\theta_{2})Y_{1}(\theta_{2}) - {2a} \chi J_{0}(\sqrt{\epsilon}\theta_{2})Y_{2}(\theta_{2}) \\
    &\;\;- 2 \sqrt{\epsilon} {\tilde{s}} J_{1}(\sqrt{\epsilon}\theta_{2})Y_{0}(\theta_{2}) + {2 \epsilon a} \chi J_{2}(\sqrt{\epsilon}\theta_{2})Y_{0}(\theta_{2}), \\
    C_{2} &= A_{2}, \\
    D_{1} &= -2{\tilde{s}} J_{0}(\sqrt{\epsilon}\theta_{2})J_{1}(\theta_{2}) + {2a} \chi J_{0}(\sqrt{\epsilon}\theta_{2})J_{2}(\theta_{2}) \\
    &\;\;+ 2 \sqrt{\epsilon} {\tilde{s}} J_{1}(\sqrt{\epsilon}\theta_{2})J_{0}(\theta_{2}) - {2 \epsilon a} \chi J_{2}(\sqrt{\epsilon}\theta_{2})J_{0}(\theta_{2}), \\
    D_{2} &= B_{2}, \\
\end{split}
\end{equation}
where $\theta_{2} \equiv 2\chi a/{\tilde{s}}$.
The analytic solution (Eq.~\ref{eq:omega_2d_approx}) for $s=1$\,cm is plotted in Fig.~\ref{fig:omega_2d} as a function of dielectric properties, size $2a$ and relative permittivity $\epsilon$.
This solution is compared with the solution obtained numerically using the square-shaped unit cell in Fig.~\ref{fig:geometry_2d}.
The geometric approximation is validated by reasonable agreement between the two approaches.
For $2a=0.2$\,cm and $\epsilon=10$, for instance, the approximate solution gives $\omega/2\pi = 19.6$\,GHz, which is consistent with the numerical computation within 3\%.

\begin{figure}
    \centering
    \includegraphics[width=\linewidth]{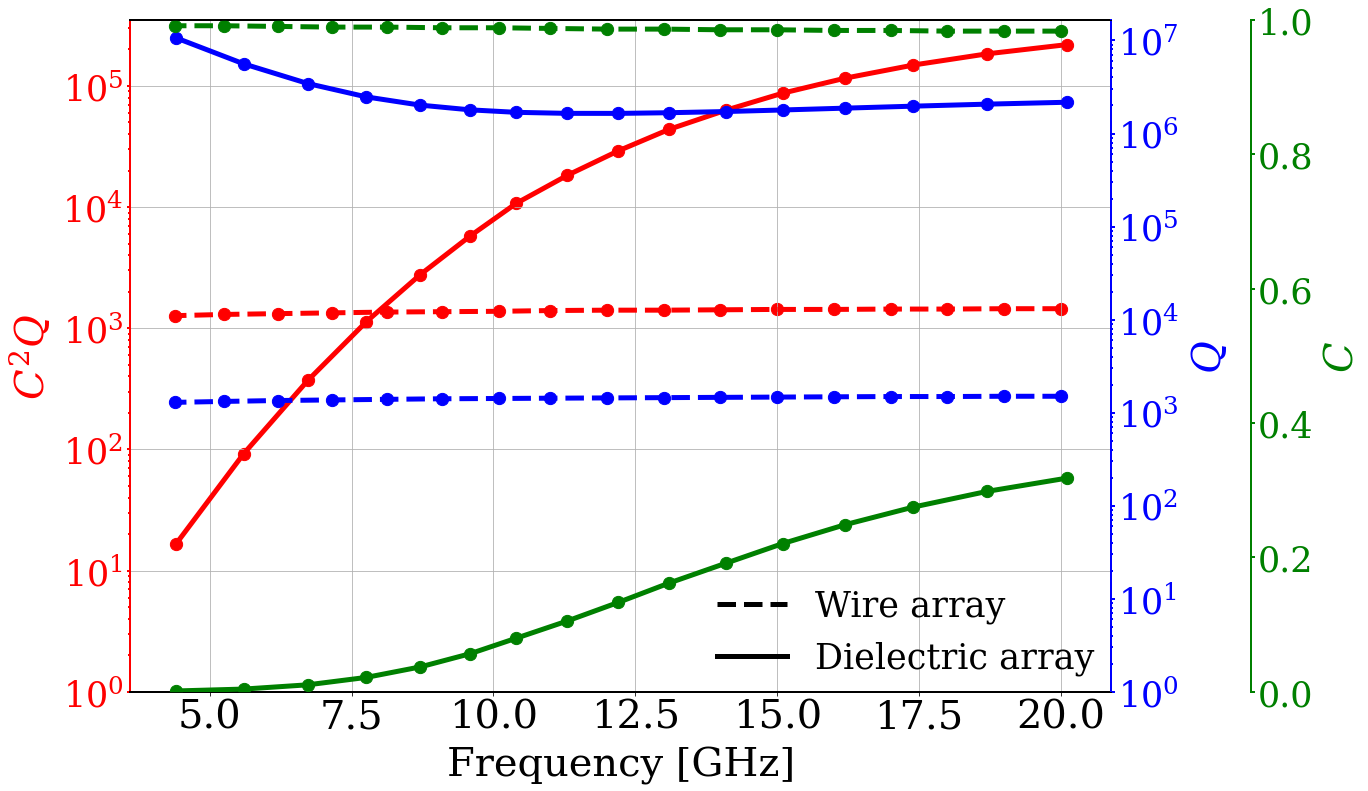}
    \caption{Performance comparison between wire (dashed lines) and dielectric (solid lines) arrays as a function of resonant frequency.
    The area below the solid lines, $\int_{\Delta f} (C^2Q) df$, corresponds to our FOM.}
    \label{fig:performance_comparison_inf}
\end{figure}

For performance evaluation, we ran a simulation to compute the cavity-relevant properties -- quality factor and form factor. 
A 2-D lattice structure was modeled as an infinite tessellation of the square unit cells in Fig.~\ref{fig:wire_vs_dielectric} subjected to periodic boundary conditions.
We assumed $\epsilon = 10$ and $\tan\delta=10^{-6}$ for the relative permittivity and dissipation factor of the dielectrics.
To optimize the dimensions of the dielectric, we use the quantity $\int_{\Delta f} (C^2Q) df$ from Ref.~\cite{paper:wheel} as a figure of merit (FOM), which reflects both experimental sensitivity (Eq.~\ref{eq:scan_rate}) and frequency tunability.
In this work, the unit cell was designed to have an initial resonant frequency of 20\,GHz.
For a fixed dimension of the dielectric rod, the resonant frequency decreases with increase in the lattice constant, i.e., with increasing interspace.
Figure~\ref{fig:performance_comparison_inf} is a plot of the performance of the dielectric photonic crystal, which is compared with that of the wire metamaterial.
The latter is also modeled in a similar manner for 20\,GHz, assuming a wire radius of 1\,mm used in Ref.~\cite{paper:wire_metamatl} and an electrical conductivity ($\sigma$) of $6.0\times10^8$\,S/m.
The value quoted for conductivity corresponds to an effective value for commercially available oxygen-free copper, measured at our laboratory in the cryogenic and GHz region.
We find that, mainly due to the substantial improvement in the quality factor, the dielectric photonic crystal is highly profitable in the upper octave band.
The FOM values (area under the red curves in Fig.~\ref{fig:performance_comparison_inf}) between 10 and 20 GHz are $1.4\times10^4$\,GHz and $9.5\times10^5$\,GHz for the wire and dielectric arrays, respectively.

For a more realistic comparison, a three-dimensional (3-D) simulation study was conducted.
First, we reproduced the numerical results of the prototype resonator (100\,mm$\times$100\,mm$\times$100\,mm) that was used to demonstrate the wire metamaterial (WM) haloscope in Ref.~\cite{paper:wire_metamatl}, where a 10$\times$10 array of 2\,mm-thick metal wires generated a resonant frequency of 11.42\,GHz.
Second, we designed a photonic crystal (PC) resonator using a 6$\times$6 array of 7.45\,mm-thick dielectric ($\epsilon=10$) rods within a cube with the same outer boundary.
For performance testing, numerical values of key parameters were obtained and a quantity $V^2C^2Q$ was computed for various values of $\sigma$ and $\tan\delta$, which are summarized in Tables~\ref{tab:parameters} and \ref{tab:performance_comparison}, respectively.
A multiple-cell (MC) cavity~\cite{paper:multiple_cell} with 16 cells was also considered for reference.

\begin{table}[h]
    \centering
    \begin{tabular*}{\columnwidth}{@{\extracolsep{\fill}}c|@{}c@{}c@{}c}
    \hline
         &  ~~~~~WM~~~~~ & ~~~~~PC~~~~~ & ~~~~~MC~~~~~ \\
    \hline\hline
    $\nu$ (GHz) & 11.42 & 11.33 & 11.31 \\
    ~~~$V$ (10$^{-3}$\,m$^{3}$)~~~ & 0.97 & 1.00 & 0.89\\
    $C$ & 0.60 & 0.24 & 0.51 \\
    $Q$ & 2080 & 15000 & 6320\\
    \hline
    \end{tabular*}
    \caption{Numerically calculated values of experimental parameters for different cavity designs. 
    For $Q$, $\sigma=1.5\times10^7$\,S/m (Ref.~\cite{paper:wire_metamatl}) and $\tan\delta=10^{-4}$ were assumed.}
    \label{tab:parameters}
\end{table}

\begin{table}[h]
    \begin{center}
    \begin{tabular*}{\columnwidth}{@{\extracolsep{\fill}}c@{}c|@{}c@{}c@{}c}
    \hline
    &     & \multirow{2}{*}{WM} & \multirow{2}{*}{PC} & \multirow{2}{*}{MC} \\
    $\sigma$ (S/m) & $\tan\delta$ & \\
    \hline\hline
    $6\times10^{7}$  & $1\times10^{-4}$ & $1.4\times10^{-3}$ & $1.0\times10^{-3}$ & $2.6\times10^{-3}$ \\
    $6\times10^{8}$  & $1\times10^{-6}$ & $4.5\times10^{-3}$ & $1.9\times10^{-2}$ & $8.3\times10^{-3}$ \\
    $1\times10^{10}$ & $1\times10^{-8}$ & $1.8\times10^{-2}$ & $9.0\times10^{-2}$ & $3.4\times10^{-2}$ \\
    \hline
    \end{tabular*}
    \caption{Computed values of the quantity $V^2C^2Q$ in units of m$^{-6}$ for performance comparison among those cavity designs with various combinations of $\sigma$ and $\tan\delta$.
    The first two rows represents commercially available copper and aluminum oxide at room and cryogenic temperatures, respectively, while the third row assumes superconductor and high-quality dielectric material~\cite{paper:sapphire_loss}.}
    \label{tab:performance_comparison}
    \end{center}
\end{table}

\vspace{0.75cm}
\section{Frequency tuning mechanism}
\label{sec:tuning_mechanism}

\begin{figure}
    \centering
    \includegraphics[width=0.9\linewidth]{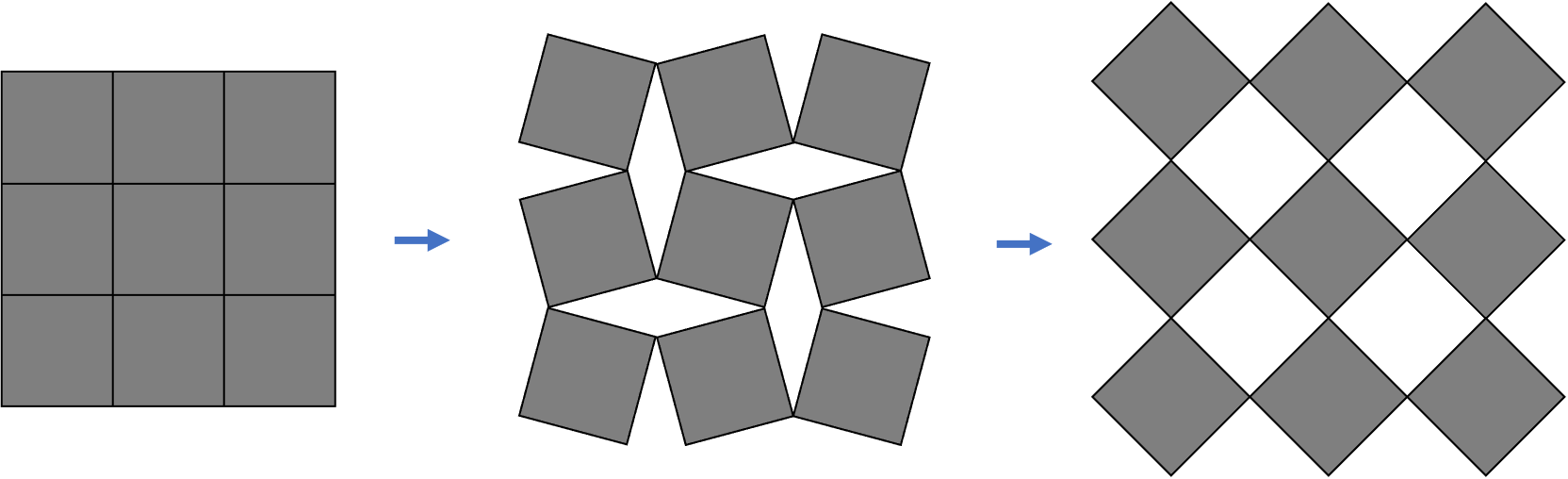}
    \caption{Auxetic behavior of rotating rigid squares. 
    With the vertices connected together, the relative rotation of the unit cells forms more open or closed structures.}
    \label{fig:auxetic_square}
\end{figure}

Similar to the plasma frequency of the wire metamaterials discussed in Ref.~\cite{paper:plasma_haloscope}, the resonant frequency of the dielectric photonic crystal can be tuned by varying the space between adjacent rods.
An ideal tuning mechanism requires isotropic expansion or contraction of the array structure in two dimensions.
Inspired by auxetics--structures or materials exhibiting a negative Poisson's ratio--we considered 2-D arrangements of rigid polygons~\cite{paper:rotating_squares}.
The polygons are connected together at their vertices in such manner that the overall structure expands when stretched and contracts when compressed, as illustrated in Fig.~\ref{fig:auxetic_square}. 
In particular, the regular tessellations present a highly symmetric pattern so that the relative rotation of the unit cells isotropically deforms the structure.
Such an auxetic-inspired deformable structure was employed for our frequency tuning.

\begin{figure}
    \centering
    \includegraphics[width=\linewidth]{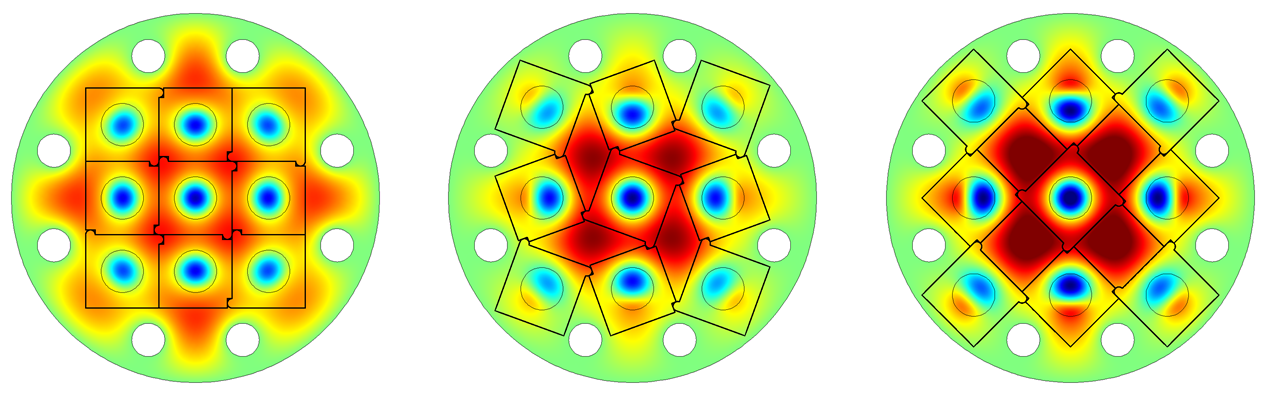}
    \caption{Two-dimensional design of a photonic crystal haloscope with a 3$\times$3 square array (black square lines) in a conducting cylindrical cavity.
    A dielectric rod is placed in the center of each square.
    The array structure deforms in the same pattern shown in Fig.~\ref{fig:auxetic_square} and the corresponding E-field distributions are displayed using the same color scale shown in Fig.~\ref{fig:wire_vs_dielectric}.
    Eight metal poles (white circles) are deployed to prevent field localization.}
    \label{fig:frequency_tuning}
\end{figure}

For a realistic design applicable to axion haloscopes, we performed a 2-D simulation study.
We modeled a cylindrical cavity of a normal conductor with a tuning structure of 3$\times$3 square blocks (forming a regular tessellation) with a dielectric rod fixed at the center of each block.
Depending on the position within the array structure, each block has a different number of ears on its corners where a hinge is introduced to join the neighboring blocks for rotational deployment.
The expansion or contraction of the structure can be achieved by rotating the center block in the same manner shown in Fig.~\ref{fig:auxetic_square}.
It was noticeable that all the tuning rods moved only radially with respect to the center.
The size and spacing of the rods are optimized to yield the highest FOM around 10\,GHz, and the auxetic structure is designed accordingly.
A set of conducting poles were strategically deployed inside the cylinder to prevent field localization and to form a well-defined resonant mode over the entire tuning range.
Figure~\ref{fig:frequency_tuning} illustrates how the pattern and electric field change for various rotation angles. 
The resonant frequency of our desired mode extends from 9.5 to 10.8\,GHz,  giving $\Delta f/f$ of $\sim15\%$. 

\section{Cavity design demonstration}
\begin{figure}
    \centering
    \includegraphics[width=0.45\linewidth]{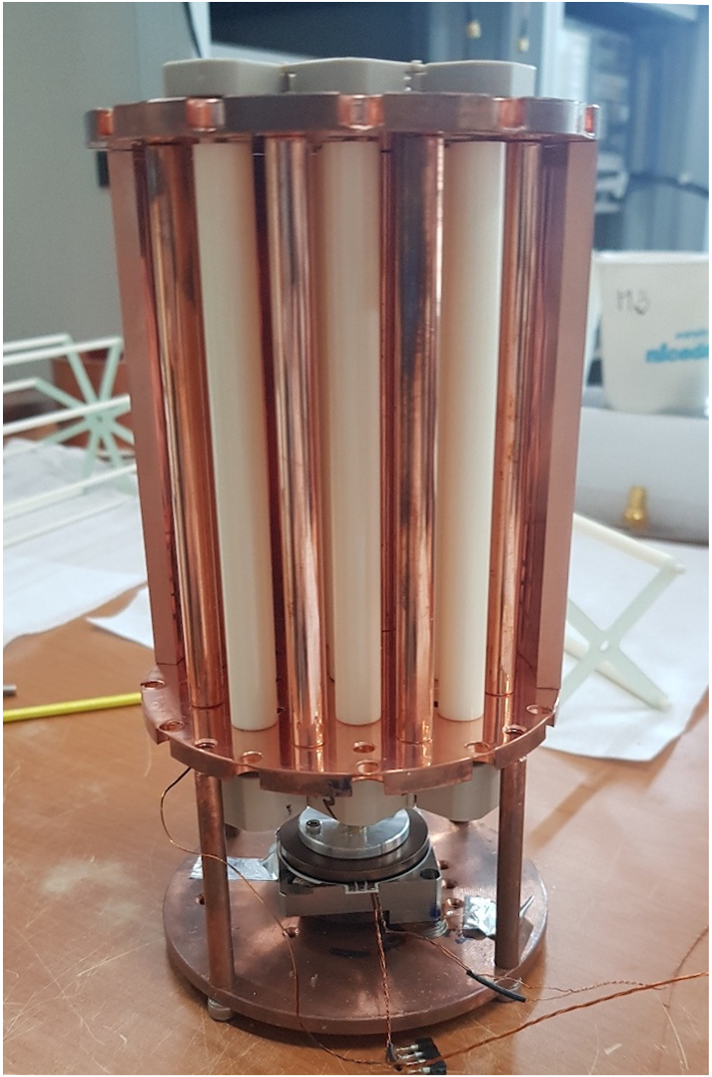}
    \caption{Photo of the photonic crystal cavity and frequency tuning system for demonstration.}
    \label{fig:skeleton_photo}
\end{figure}

The experimental feasibility of this photonic crystal cavity with an auxetic structure for tunable axion haloscopes was tested by demonstrating the cavity design depicted in Section~\ref{sec:tuning_mechanism}.
Made of oxygen-free high conductivity copper with 99.99\% nominal purity, the cavity consisted of a 5\,mm-thick cylindrical wall with 78\,mm inner diameter and 100\,mm height, along with two 5\,mm-thick endcaps.
The tuning rods, made of commercially available high-purity aluminum oxide (99.7\%) for large field distortion ($\epsilon_r=9.7$) and low dissipation ($\tan\delta\sim10^{-6}$ at 4\,K), were 9\,mm in diameter and long enough to be extended out of the cavity and be gripped at both ends by a pair of auxetic structures that were fabricated from polyether ether ketone. 
Both ends of each rod was machined to be thin (2\,mm in diameter) to prevent radiation leaks through the openings introduced in the cavity endcaps.
The auxetic structure was designed in the same way as discussed in Section~\ref{sec:tuning_mechanism} with a unit square block having a dimension of 15.5$\times$15.5$\times$10.0\,mm$^3$.
A piezoelectric rotary actuator was attached to the center block to induce the rotational motion of the auxetic structure.
Since the tuning rods move only radially as mentioned earlier, the endcaps have slotted openings along the radial direction to guide their movement.
Four pairs of 7\,mm-thick copper poles were placed near the inner edge of the cavity as shown in Fig.~\ref{fig:frequency_tuning}.
The overall structure of this demonstration cavity design can be seen in Fig.~\ref{fig:skeleton_photo}.

\begin{figure}[b]
    \centering
    \includegraphics[width=0.9\linewidth]{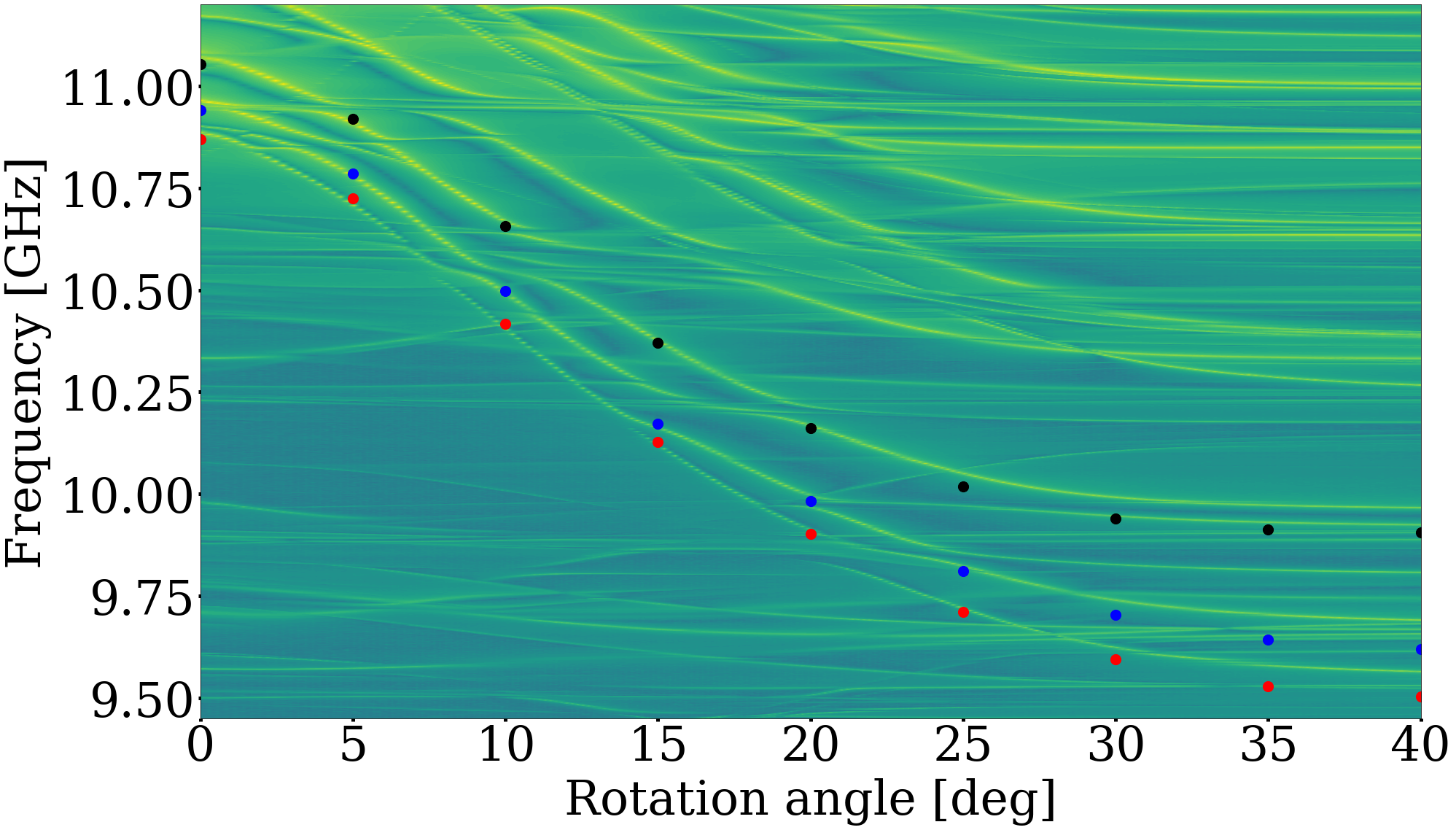}
    \caption{Mode map of the prototype cavity depicted in the text.
    The resonant frequencies of the first three TM modes are in good agreement with simulation represented by colored dots.
    Our desired mode corresponds to the lowest curve (with overlapping red dots) that spans the frequency from 10.9 to 9.6\,GHz. 
    }
    \label{fig:mode_map}
\end{figure}

The demonstration was performed at a cryogenic temperature.
The cavity assembly was installed in a vacuum chamber immersed in a liquid helium reservoir.
The cavity and dielectric rods were cooled to 4.5\,K by helium exchange gas injected into the chamber.
The cavity properties were measured through a transmission signal between a pair of weakly coupled coaxial antennae.
The frequency tuning was achieved by rotating the center block of the auxetic structure using the piezoelectric rotator.
Figure~\ref{fig:mode_map} shows the frequency map of the cavity design over a full rotation angle (40$^{\circ}$). 
The desired mode, which corresponds to the lowest curve, had a frequency span of about 1.3\,GHz from 10.9 to 9.6 GHz, which agrees well with the simulation.
Some mode crossings are observed, appearing as discontinuities along the curve, notably near 10.5 and 9.8\,GHz.
Since mode mixing gives rise to a considerable reduction in the form factor, those local regions are typically not usable in axion searches, eventually limiting the dynamic tuning range.

\begin{figure}
    \centering
    \includegraphics[width=\linewidth]{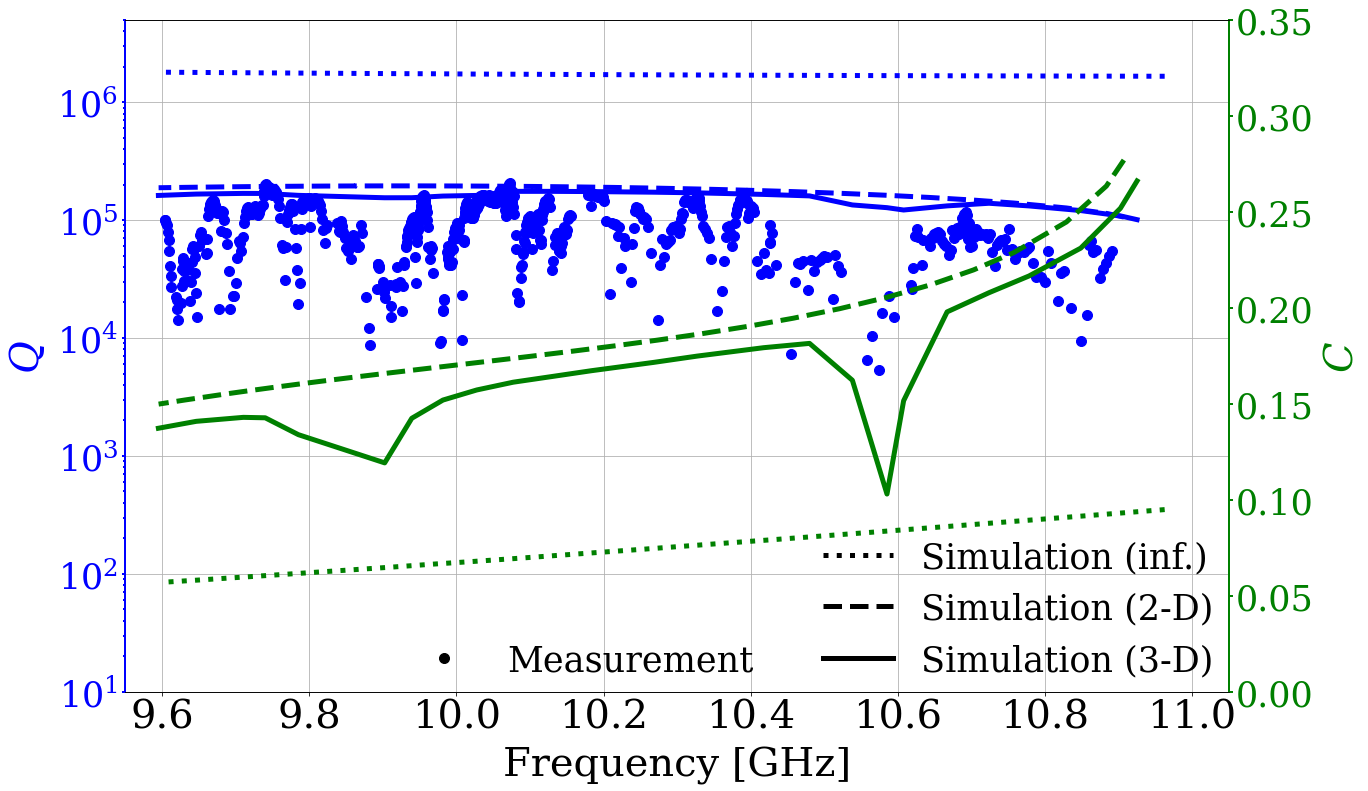}
    \caption{Measured quality factors of the demonstration cavity, compared to simulation results obtained from modeling in different dimensions assuming $\tan\delta=5\times10^{-6}$.
    The values for an infinitely periodic array labeled `Simulation (inf)' are taken from Fig.~\ref{fig:performance_comparison_inf}. 
    The simulated values of form factor are also overlapped.}
    \label{fig:Q_C_comparison}
\end{figure}

The cavity quality factor was measured during the tuning process and compared to sparsely performed simulations results in Fig.~\ref{fig:Q_C_comparison}.
The measurements reach the highest value of $\sim 210,000$ near 10.1\,GHz and slowly descends in both lower and higher frequency regions.
From these values, the dissipation factor of the dielectric rods was estimated to be $\tan\delta\sim5\times 10^{-6}$, which is reasonable for nominal aluminum oxide.
The intermittent drops in quality factor could be attributed to longitudinal geometric symmetry breaking due to mechanical misalignment, e.g., gap or tilt, of the tuning rods resulting in increased mode complexity~\cite{paper:mode_crossing}.
The simulated form factors were also overlapped in the figure where two notable drops correspond to the major mode-crossing regions mentioned above.
This verifies the experimental feasibility of the new cavity design and the newly invented tuning mechanism.
Furthermore, we observed 5\% increase in the quality factor at magnetic fields greater than 1\,T, which is likely due to the similar magnetic property reported in Ref.~\cite{paper:QUAX_highQ}. 
The enhancement will be more significant with higher-purity dielectric material at mK temperatures~\cite{paper:highQ_mK}.
This concludes that the photonic crystal haloscope is clearly advantageous in the search for high-mass axions with high performance.

\section{Conclusion}
We proposed a new cavity design of photonic crystal suitable for dark matter axion searches in the high microwave region, which is not practically accessible using conventional cavity haloscopes.
The design features a lattice structure of high-quality dielectric material, which substantially improved the quality factors.
The EM solutions for our desired mode were obtained using an analytical approach to a good approximation and the major characteristics were examined via numerical calculations.
We showed that dielectric metamaterials are superior to metallic (plasmonic) metamaterials in terms of search performance.
The auxetic behavior from rotating squares of regular tessellation allows a two-dimensional isotropic movement of the dielectric array, which serves as a novel frequency tuning mechanism. 
An experimental demonstration verified that this new cavity haloscope design offers a highly efficient approach for high-mass axion searches.

\section*{acknowledge}
This work was supported by the Institute for Basic Science (IBS-R017-D1-2022-a00).
We thank Y. Jeong for providing the novel idea for frequency tuning based on kirigami.



\begin{thebibliography}{99}
\bibitem{paper:axion1} R. D. Peccei and H. R. Quinn, {\it CP Conservation in the Presence of Pseudoparticles}, Phys. Rev. Lett. {\bf 38}, 1440 (1977); R. D. Peccei and H. R. Quinn, {\it Constraints imposed by CP conservation in the presence of pseudoparticles}, Phys. Rev. D {\bf 16}, 1791 (1977).
\bibitem{paper:axion2} S. Weinberg, {\it A new light boson?}, Phys. Rev. Lett. {\bf 40}, 223 (1978); F. Wilczek, {\it Problem of strong P and T invariance in the presence of instantons}, Phys. Rev. Lett. {\bf 40}, 279 (1978).
\bibitem{paper:CDM} J. Preskill, M. B. Wise and F. Wilczek, {\it Cosmology of the invisible axion}, Phys. Lett. B {\bf 120}, 127 (1983); L. F. Abbott and P. Sikivie, {\it A cosmological bound on the invisible axion}, Phys. Lett. B {\bf 120}, 133 (1983); M. Dine and W. Fischler, {\it The not-so-harmless axion}, Phys. Lett. B {\bf 120}, 137 (1983).
\bibitem{paper:sikivie} P. Sikivie, {\it Experimental Tests of the ``Invisible" Axion}, Phys. Rev. Lett. {\bf 51}, 1415 (1983).
\bibitem{paper:detection_rate}D. Kim {\it et al.}, {\it Revisiting the detection rate for axion haloscopes}, J. Cosmol. Astropart. Phys. {\bf 03}, 066 (2020).

\bibitem{thesis:ADMX_multicav} D. S. Kinion, {\it First results from a multiple microwave cavity search for dark matter axions}, Ph.D. Thesis, University of California, Davis (2001).
\bibitem{paper:multiple_cavity} J. Jeong {\it et al.}, {\it Phase-matching of multiple-cavity detectors for dark matter axion search}, Astropart. Phys. {\bf 97} 33 (2017).
\bibitem{paper:CAPP-9T} J. Jeong {\it et al.}, {\it Search for Invisible Axion Dark Matter with a Multiple-Cell Haloscope}, Phys. Rev. Lett. {\bf 125}, 221302 (2020).
\bibitem{paper:RADES} A. Á. Melcón, S. A. Cuendis, C. Cogollos, {\it et al.}, {\it Scalable haloscopes for axion dark matter detection in the 30 $\mu$eV range with RADES}, J. High Energ. Phys. {\bf 2020}, 84 (2020).
\bibitem{paper:supermode} B. T. McAllister, G. Flower, L. E. Tobar, and M. E. Tobar, {\it Tunable Supermode Dielectric Resonators for Axion Dark-Matter Haloscopes}, Phys. Rev. Applied {\bf 9}, 014028 (2018).
\bibitem{paper:wheel} J. Kim {\it et al.}, {\it Exploiting higher-order resonant modes for axion haloscopes}, J. Phys. G: Nucl. Part. Phys. {\it 47}, 035203 (2020).
\bibitem{paper:DBAS} A. P. Quiskamp, B. T. McAllister, G. Rybka, and M. E. Tobar, {\it Dielectric Boosted Axion Haloscope Sensitivity In Cylindrical Azimuthally Varying Transverse Magnetic Resonant Modes}, Phys. Rev. Applied {\bf 14}, 044051  (2020).
\bibitem{paper:SC_QUAX} D. Alesini, C. Braggio, G. Carugno, N. Crescini, D. D’Agostino, D. Di Gioacchino et al. (QUAX Collaboration), {\it Realization of a high quality factor resonator with hollow dielectric cylinders for axion searches}, Nucl. Instrum. Methods Phys. Res., Sect. A {\bf 985}, 164641 (2021).
\bibitem{paper:highQ_QUAX} D. Alesini(Frascati) {\it et al.} (QUAX Collaboration), {\it Realization of a high quality factor resonator with hollow dielectric cylinders for axion searches}, Nucl. Instrum. Meth. A {\bf 985}, 164641 (2021).
\bibitem{paper:SC_CAPP} D. Ahn et al., Superconducting cavity in a high magnetic field, arXiv:2002.08769.
\bibitem{paper:MADMAX} A. Caldwell {\it et al.} (MADMAX Working Group), {\it Dielectric haloscopes: a new way to detect axion dark matter}, Phys. Rev. Lett. {\bf 118}, 091801 (2017).
\bibitem{paper:plasma_haloscope} M. Lawson, A. J. Millar, M. Pancaldi, E. Vitagliano, and F. Wilczek, {\it Tunable Axion Plasma Haloscopes}, Phys. Rev. Lett. {\bf 123}, 141802 (2019).

\bibitem{tool:comsol} COMSOL Multiphysics$\textsuperscript{\textregistered}$ v. 5.2. www.comsol.com. COMSOL AB, Stockholm, Sweden.
\bibitem{paper:Dirac_cone} C. T. Chan, Z. H.  Hang and X. Huang, {\it Dirac Dispersion in Two-Dimensional Photonic Crystals}, Adv. OptoEletron, {\bf 2012}, 313984 (2012); Y. Li, C. T. Chan and E. Mazur, {\it Dirac-like cone-based electromagnetic zero-index metamaterials}, Light: Sci. Appl. {\bf 10}, 203 (2021); J. Luo and Y. Lai, {\it Hermitian and Non-Hermitian Dirac-Like Cones in Photonic and Phononic Structures}, Front. Phys. {\bf 10}, 845624 (2022).

\bibitem{paper:wire_metamatl} R. Balafendiev, C. Simovski, P. Belov, and A. J. Millar, {\it Wire metamaterial filled metallic resonators}, Phys. Rev. B {\bf 106}, 075106 (2022).
\bibitem{paper:multiple_cell} J. Jeong {\it et al.}, {\it Concept of multiple-cell cavity for axion dark matter search}, Phys. Lett. B {\bf 777}, 412 (2018).

\bibitem{paper:sapphire_loss} J. Krupka, K. Derzakowski, M. Tobar, J. Hartnett, and  R. G. Geyer, {\it Complex permittivity of some ultralow loss dielectric crystals at cryogenic temperatures} Measurement Science and Technology, {\bf 10(5)}, 387 (1999).

\bibitem{paper:rotating_squares} J. N. Grima and K. E. Evans, {\it Auxetic behavior from rotating squares}, J. Mater. Sci. Lett. {\bf 19}, 1563–1565 (2000); J. N. Grima, A. Alderson and K. E. Evans, {\it Auxetic behaviour from rotating rigid units}, Phys. Status Solidi B {\bf 242}, 561 (2005).

\bibitem{paper:QUAX_highQ} R. Di Vora {\it et al.}, {\it A high-Q microwave dielectric resonator for axion dark matter haloscopes}, arXiv:2201.04223 (2022).
\bibitem{paper:highQ_mK} D. L. Creedon {\it et al.}, {\it High $Q$-factor sapphire whispering gallery mode microwave resonator at single photon energies and millikelvin temperatures} Appl. Phys. Lett. {\bf 98}, 222903 (2011).
\bibitem{paper:mode_crossing} I. Stern, G. Carosi, N.S. Sullivan, and D.B. Tanner, {\it Avoided Mode Crossings in Cylindrical Microwave Cavities}, Phys. Rev. Applied {\bf 12}, 044016 (2019).

\end{thebibliography}
\end{document}